\documentstyle[aps,preprint,draft,epsf]{revtex}
\title{Evolution of Cosmological Perturbation in Reheating Phase 
of the Universe}
\date{September 12, 1996}
\author{Yasusada Nambu and Atsushi Taruya}
\address{Department of Physics, Nagoya University \\ 
        Chikusa, Nagoya 464-01, Japan}
\begin{document}
\maketitle
\begin{abstract}
The evolution of the cosmological perturbation during the 
oscillatory stage of the scalar field is 
investigated. For the power law potential of the inflaton field,  
the evolution equation of the Mukhanov's 
gauge invariant variable is reduced to the Mathieu equation and 
the density perturbation grows by the parametric resonance.
\end{abstract}
%%%%%%%
\def\mpl{m_{pl}}
\def\tphi{\tilde\phi}
\def\tq{\tilde Q}
%%%%%%%%%%%%%%%%%%%%%%%%%%%%%%%%%%%%%%%%%
%\section{introduction}
%\begin{itemize}
%	\item  Evolution of cosmological perturbation in coherent 
%	oscillationary phase of the scalar field(reheating).
%
%	\item  Use of Mukhanov's gauge invariant quantity
%
%	\item  Reduction of perturbation equation to Mathieu equation
%	
%	\item Parametric resonance and growth of density perturbation
%\end{itemize}

Recently, the importance of the parametric resonance is recognized in 
the reheating phase of the inflationary model\cite{sht,kof}. Due to the coherent 
oscillation of the background inflaton field, the fluctuation of the 
non-linearly coupled boson field or the fluctuation of 
self interacting inflaton field itself are amplified and the 
catastrophic particle creation occurs. This effect drastically 
changes the scenario of the reheating considered so far. When one 
 pay attention to the evolution of the metric 
perturbation during the coherent oscillatory phase of the scalar 
field dominated universe, a naive question arises:does the metric 
perturbation undergo a influence of the parametric resonance by  
the background oscillation of the scalar field? 

It is well known that the cosmological perturbation of the scalar field 
in the oscillatory stage has the problematic aspect\cite{ks}. If one try to 
write down the evolution equation for the density contrast or the 
Newtonian potential, the coefficient of the equation becomes singular 
periodically in time due to the background oscillation of the scalar 
field. This behavior is not real. It appears through the reduction of 
the constrained system to the second order differential equation. But 
it makes difficult to understand the behavior of the metric 
perturbation in the oscillatory phase.

In this paper, we consider the evolution of the metric perturbation 
using the gauge invariant variable introduced by Mukhanov. The 
evolution equation for this variable has no singular behavior and is 
suitable to apply the oscillatory phase of the scalar field. 
%%%%%%%%%%%%%%%%%%%%%%%%%%%%%%%%%%%%%%%%%%
%\section{basic equation}
%\subsection{background equation}
We treat a spatially flat FRW universe with a minimally 
coupled scalar field with a potential
%%%
\begin{equation}
 V(\phi)=\frac{\lambda}{4}\phi_{0}^{4}
  \left(\frac{\phi}{\phi_{0}}\right)^{2n},~~(n=1,2,\ldots)
\end{equation}
%%%
The background equations are
%%%
\begin{eqnarray}
 H^{2}&=&\frac{\kappa}{3}(\frac{1}{2}\dot\phi^{2}+V(\phi)), \\
 \dot H&=&-\frac{\kappa}{2}\dot\phi^{2}, \\
 \ddot\phi&+&3H\dot\phi+V_{\phi}=0,
\end{eqnarray}
%%%
where $\kappa=8\pi G$.
The scalar field oscillates if the condition $\phi\ll\mpl$ is 
satisfied. In such a situation, we cannot have the exact solution, but 
using the time averaged potential energy and 
the kinetic energy have the relation 
%%%
\begin{equation}
 <\dot\phi^{2}>=2n<V(\phi)>, 
\end{equation}
%%%
the scale factor and the Hubble parameter can be approximately expressed as
%%%
\begin{equation}
 a(t)\approx\left(\frac{t}{t_{0}}\right)^{\frac{n+1}{3n}},
 ~~~H\approx\frac{n+1}{3n}\frac{1}{t}.
\end{equation}
%%%
We define new variables for the scalar field and the time:
%%%
\begin{eqnarray}
  \eta&=&nt_{0}a^{\frac{3}{n+1}}=nt_{0}\left(\frac{t}{t_{0}}\right)^{1/n}, \\
  \phi(t)&=&\phi_{0}a^{-\frac{3}{n+1}}\tphi,
\end{eqnarray}
%%%
where $\tphi=1$ at $t=t_{0}$ and $\phi_{0}$ is a initial value of the scalar 
field($\phi_{0}\ll\mpl$). Using the new variables, the evolution 
equation of the scalar field becomes  
%%%
\begin{equation}
 \tphi_{\eta\eta}+m^{2}n\tphi^{2n-1}=0,~~~
     m^{2}=\frac{\lambda}{2}\phi_{0}^{2}. \label{sc}
\end{equation}
%%%
For $n=1$(massive scalar field), $\tphi=\cos(m\eta)$. For $n=2$, 
$\tphi=cn(\sqrt{\lambda}\phi_{0}\eta;\frac{1}{\sqrt{2}})$ where $cn$ 
is a elliptic function.

%%%%%%%%%%%%%%%%%%%%%%%%%
%\subsection{perturbation equation}
We use the gauge invariant variables to treat the perturbation whose 
wavelength is larger than the horizon scale. We found that the most 
convenient variable is 
%%%
\begin{equation}
 Q=\delta\phi-\frac{\dot\phi}{H}\psi=\delta\phi^{(g)}-\frac{\dot\phi}{H}\Psi,
\end{equation}
%%%
where $\psi$ is the perturbation of the three curvature, $\delta\phi^{(g)}$ 
is the gauge invariant variable for the scalar 
field perturbation and $\Psi$ is the gauge invariant Newtonian 
potential. For the zero curvature slice, $Q$ represents the fluctuation of the 
scalar field. This variable $Q$ was first introduced by 
Mukhanov\cite{muk}. As already mentioned, the coefficient of 
the evolution equation for the Newtonian 
potential or the gauge invariant density contrast becomes singular  
 because of the background oscillation of the scalar field. 
 But the evolution equation for $Q$ 
does not have such a singular behavior:
%%%
\begin{equation}
\ddot Q+3H\dot Q+\left[V_{\phi\phi}+\left(\frac{k}{a}\right)^{2}+
  2\left(\frac{\dot H}{H}+3H\right)^{\cdot}\right]Q=0. \label{muk}
\end{equation}
%%%
The Newtonian potential and the variable $Q$ is connected by the 
relation
%%%
\begin{equation}
 -\frac{k^{2}}{a^{2}}\Psi=\frac{\kappa}{2}\frac{\dot\phi^{2}}{H}
     \left(\frac{H}{\dot\phi}Q\right)^{\cdot}, \label{pot}
\end{equation}
%%%
and the gauge invariant density contrast $\Delta$ which is equal to 
$\left(\frac{\delta\rho}{\rho}\right)$ on the co-moving time slice is
%%%
\begin{equation}
 \Delta= \frac{\kappa}{3}
 \frac{\dot\phi^{2}}{H^{3}}\left(\frac{H}{\dot\phi}Q\right)^{\cdot}. 
 \label{delta}
\end{equation}
%%%%%%%%%%%%%%%%%%%%%%%%%%%%%%%%%%%%%%%%%%%%%%%%%%%%%%%%%%
%\section{behavior of gauge invariant variable}
To treat the eq.(\ref{muk}) more tractable, we change the variable
%%%
\begin{equation}
 Q=a^{-\frac{3}{n+1}}\tq,~~~
 \eta=nt_{0}\left(\frac{t}{t_{0}}\right)^{1/n},
\end{equation}
%%%
Then
%%%
\begin{equation}
 \tq_{\eta\eta}+a^{\frac{6(n-1)}{n+1}}\left[
  V_{\phi\phi}+\left(\frac{k}{a}\right)^{2}+2\left(\frac{\dot 
  H}{H}+3H\right)^{\cdot}-\frac{9n}{(n+1)^{2}}H^{2}-\frac{3}{n+1}\dot 
  H\right]\tq=0. \label{muk2}
\end{equation}
%%%
Using the background equation, we can estimate the time dependence of 
the each terms in this equation:
%%%
\begin{eqnarray}
&&V_{\phi\phi}=n(2n-1)m^{2}a^{-\frac{6(n-1)}{n+1}}\tphi^{2n-2}\sim 
 O\left(\left(\frac{\eta}{t_{0}}\right)^{2-2n}\right), \nonumber \\
&&\left(\frac{\dot H}{H}+3H\right)^{\cdot}=
 6Hn^{n}\left(\frac{\eta}{t_{0}}\right)^{1-n}\tphi^{2n-1}\tphi_{\eta}
 \sim O\left(\left(\frac{\eta}{t_{0}}\right)^{1-2n}\right), \nonumber \\
&&H^{2},~\dot H\sim O\left(\left(\frac{\eta}{t_{0}}\right)^{-2n}\right). \nonumber
\end{eqnarray}
%%%
As we are considering the situation $\eta\geq t_{0}$, we can 
neglect the $H^{2}, \dot H$ terms in eq.(\ref{muk2}) because they are 
higher order in powers of $\left(1/\eta\right)$ compared to other terms. Our basic 
equation for the gauge invariant variable becomes
%%%
\begin{equation}
 \tq_{\eta\eta}+\left[n(2n-1)m^{2}\tphi^{2n-2}
  +k^{2}a^{\frac{4(n-2)}{n+1}}
  +4n(n+1)\frac{1}{\eta}\tphi^{2n-1}\tphi_{\eta}\right]\tq=0.
  \label{muk3}
\end{equation}
%%%
$\tphi$ is the solution of eq.(\ref{sc}).
%%%%%%%%%%%%%%%%%%%%%%%%%%%%
%\subsection{$n=1$ case}
We first consider $n=1$ case(massive scalar). The background scalar 
field is sinusoidal and given by
\begin{equation}
 \tphi=\cos(m\eta). 
\end{equation}
Eq.(\ref{muk3}) becomes
%%%
\begin{equation}
 \tq_{\eta\eta}+\left[m^{2}+\left(\frac{k}{a}\right)^{2}-
  \frac{8}{\eta}\sin(m\eta)\cos(m\eta)\right]\tq=0.
\end{equation}
%%%
We introduce a dimensionless time variable $\tau=m\eta$,
%%%
\begin{equation}
 \tq_{\tau\tau}+\left[1+\left(\frac{k}{ma}\right)^{2}
   -\frac{4}{\tau}\sin(2\tau)\right]\tq=0, \label{ma1}
\end{equation}
%%%
where $a\propto\tau^{2/3}$. This equation has the same form as 
the Mathieu equation:
%%%
\begin{equation}
 Y_{\tau\tau}+\left[A-2q\sin(2\tau)\right]Y=0. \label{math}
\end{equation}
%%%
In our case the coefficient $A, q$ are time dependent functions 
 and the relation between $A$ and $q$ is
%%%
\begin{equation}
 A=1+\left(\frac{mt_{0}}{2}\right)^{4/3}
   \left(\frac{k}{m}\right)^{2}q^{4/3}.
\end{equation}
%%%
Using the stability/instability chart of the Mathieu equation, we can know 
that the perturbation will have the effect of parametric resonance of 
the first unstable band of the Mathieu function and 
grows in time(see figure). We can derive its time evolution by solving 
eq.(\ref{ma1}) using multi-time scale method\cite{nay}. We introduce a parameter
%%%
\begin{equation}
 \epsilon=\frac{4}{\tau_{0}}.
\end{equation}
%%%
As the condition $\tau_{0}\gg 1$ is equivalent to the condition of coherent 
oscillation, $\epsilon$ is small parameter. Rewrite the 
eq.(\ref{ma1}) as
%%%
\begin{equation}
\tq_{\tau\tau}+\left[1+2\epsilon\omega_{1}-
  \epsilon\frac{\tau_{0}}{\tau}\sin(2\tau)\right]\tq=0,
  ~~(\tau\ge\tau_{0})
\end{equation}
%%%
where $\omega_{1}=\frac{1}{2\epsilon}\left(\frac{k}{ma}\right)^{2}$. 
We assume the condition $\left(\frac{k}{ma}\right)^{2}<1$ to be the 
term $2\epsilon\omega_{1}$ small. This means we consider the 
wavelength larger than the Compton length. We define 
slow time scale $\tau_{n}=\epsilon^{n}\tau$. The time derivative with 
respect to $\tau$ is replaced by
%%%
\begin{equation}
 \frac{d}{d\tau}=D_{0}+\epsilon D_{1}+\cdots,
\end{equation}
%%%
where $D_{n}=\frac{\partial}{\partial\tau_{n}}$. We expand 
%%%
\begin{equation}
 \tq=Q^{(0)}+\epsilon Q^{(1)}+\cdots.
\end{equation}
%%%
By substituting these expression to eq.(\ref{ma1}) and collect the 
terms of each order of $\epsilon$. From the $O(\epsilon^{0})$ and 
$O(\epsilon^{1})$, we have
%%%
\begin{eqnarray}
O(\epsilon^{0})&:&~~~D_{0}^{2}Q^{(0)}+Q^{(0)}=0, \label{mul1} \\
O(\epsilon^{1})&:&~~~D_{0}^{2}Q^{(1)}+Q^{(1)}=-\left(2D_{0}D_{1}Q^{(0)}+
  2\omega_{1}Q^{(0)}-\frac{\tau_{0}}{\tau}\sin(2\tau)Q^{(0)}\right). 
  \label{mul2}
\end{eqnarray}
%%%
The solution of eq.(\ref{mul1}) is
%%%
\begin{equation}
 Q^{(0)}={\cal A}(\tau_{1})e^{i\tau}+{\cal A}^{*}(\tau_{1})e^{-i\tau}. \label{e0}
\end{equation}
%%%
We substitute this to the right hand side of eq.(\ref{mul2}) and 
demand that the secular term which is proportional to $e^{i\tau}$ 
vanishes. This gives the equation for the amplitude ${\cal A}$:
%%%
\begin{equation}
 i\frac{\partial 
 {\cal A}}{\partial\tau_{1}}+\omega_{1}{\cal A}+i\frac{\tau_{0}}{4\tau}{\cal A}^{*}=0.
\end{equation}
%%%
Using the definition of $\epsilon$ and $\tau_{1}$, we have
%%%
\begin{equation}
  i\frac{\partial 
  {\cal A}}{\partial\tau}+\frac{1}{2}\left(\frac{k}{ma}\right)^{2}{\cal A}
  +\frac{i}{\tau}{\cal A}^{*}=0.
\end{equation}
%%%
Writing ${\cal A}=u+iv$($u,v$ are real), $u$ satisfies the following second 
order differential equation:
%%%
\begin{equation}
 u_{\tau\tau}+\frac{4}{3\tau}u_{\tau}+\left[
  \frac{1}{4}\left(\frac{k}{ma}\right)^{4}-\frac{2}{3\tau}\right]u=0. 
  \label{u}
\end{equation}
%%%
Using the eq.(\ref{delta}) and (\ref{e0}),
%%%
\begin{equation}
  \Delta\propto\tq_{\eta}\tphi_{\eta}-\tq\tphi_{\eta\eta}-
   \frac{1}{\eta}\tq_{\eta}\tphi=u+O(\frac{u}{\eta}).
\end{equation}
%%%
Therefore the function $u$  is equal to the 
gauge invariant density contrast $\Delta$ within the approximation we 
are using here. The solution of eq.(\ref{u}) is
%%%
\begin{equation}
 u=\tau^{-1/6}Z_{\pm 
 5/2}\left(\left(\frac{k}{a}\right)^{2}\frac{1}{mH}\right),
\end{equation}
%%%
where $Z_{\nu}$ is a Bessel function of order $\nu$.
We have the critical wavelength $\lambda_{J}=(mH)^{-1/2}$. The mode 
whose wavelength is larger than $\lambda_{J}$ can grow. If the 
wavelength is shorter than $\lambda_{J}$ initially, the wavelength is 
stretched by the cosmic expansion and its exceeds the critical length. 
We can see this behavior by using the chart of Mathieu function. The 
trajectory which started from the stable region moves to the unstable 
region.  For $k\rightarrow 0$ limit, 
%%%
\begin{eqnarray}
  \Delta&\propto&\tau^{2/3}=a,~~\tau^{-1}=a^{-3/2}, \nonumber \\
  \Psi&\propto&\hbox{constant},~~\tau^{-5/3}=\frac{H}{a}.
\end{eqnarray}
%%%
This behavior is the same as the perturbation in the dust dominated universe. 

%%%%%%%%%%%%%%%%%%%%%%%%%%%%%%%%%%%%%
%\subsection{$n\geq 2$ case}
For $n\geq 2$, the scalar filed oscillation is not sinusoidal.
We start searching the solution of the equation for $y=\tphi_{\eta}$:
%%%
\begin{equation}
 y_{\eta\eta}+n(2n-1)m^{2}\tphi^{2n-2}y=0. \label{y}
\end{equation}
%%%
Eq.(\ref{muk3}) reduces to this equation if the third and the forth 
term do not exist. We approximate the solution of eq.(\ref{y}) by 
 sinusoidal function:$y=m\sin(cm\eta)$. $c$ is a some numerical 
constant which defines the period of scalar field oscillation. Using 
the equation for $\tphi$, we have
%%%
\begin{eqnarray}
 &&n(2n-1)m^{2}\tphi^{2n-2}=-\frac{y_{\eta\eta}}{y}, \nonumber \\
 &&2n(n+1)m^{2}\tphi^{2n-1}\tphi_{\eta}
  =(\tphi_{\eta}\tphi)_{\eta\eta}, \nonumber
\end{eqnarray}
%%%
and using $y=\tphi_{\eta}=m\sin(cm\eta)$, the equation of $\tq$ becomes
%%%
\begin{equation}
 \tq_{\eta\eta}+\left[c^{2}m^{2}-\frac{4cm}{\eta}\sin(2cm\eta)
  +k^{2}a^{\frac{4(n-2)}{n+1}}\right]\tq=0.
\end{equation}
%%%
By introducing the dimensionless time variable $\tau=cm\eta$,
%%%
\begin{equation}
 \tq_{\tau\tau}+\left[1-\frac{4}{\tau}\sin(2\tau)
  +\left(\frac{k}{cm}\right)^{2}a^{\frac{4(n-2)}{n+1}}\right]\tq=0. 
  \label{ma2}
\end{equation}
%%%
This is also Mathieu equation. It is surprising that this equation contains 
the $n=1$ case if we set $c=1$. The function $A$ and $q$ are
%%%
\begin{equation}
 A=1+\alpha\tau^{\frac{4(2-n)}{3}},~~q=\frac{2}{\tau},
\end{equation}
%%%
where $\alpha=\left(\frac{k}{cm}\right)^{2}(cnt_{0}m)^{4(n-2)/3}$ and 
we have the relation
%%%
\begin{equation}
 A(q)=1+\alpha\left(\frac{q}{2}\right)^{\frac{4(2-n)}{3}}.
\end{equation}
%%%
Using the chart of Mathieu function, we find that the perturbation 
also get the effect of the parametric resonance of the first unstable 
band and grows. But as the time goes on, the trajectory moves from the 
unstable region to the stable region and the perturbation will oscillate 
with a constant amplitude. 
To investigate these behavior, we introduce slow time variable and 
derive the equation for slowly changing amplitude of $\tq$. The 
procedure is the completely same as $n=1$ case. The result is
%%%
\begin{eqnarray}
 &&\tq={\cal A}e^{i\tau}+{\cal A}^{*}e^{-i\tau},   \\
 &&i\frac{\partial {\cal A}}{\partial\tau}+
  \frac{1}{2}\left(\frac{k}{cm}\right)^{2}a^{\frac{4(n-2)}{n+1}}{\cal 
  A}
  +\frac{i}{\tau}{\cal A}^{*}=0.
\end{eqnarray}
%%%
The real part of ${\cal A}$ obeys
%%%
\begin{equation}
 u_{\tau\tau}-\frac{4}{3}(n-2)\frac{1}{\tau}u_{\tau}
  +\left[\frac{\alpha^{2}}{4}\tau^{\frac{8}{3}(n-2)}
  -\left(\frac{4}{3}n-\frac{2}{3}\right)\frac{1}{\tau^{2}}\right]u=0.
\end{equation}
%%%
The solution of this equation is
%%%
\begin{equation}
 u=\tau^{\frac{1}{6}(4n-5)}Z_{\nu}\left(\frac{3\alpha}{2(4n-5)}
  \tau^{\frac{1}{3}(4n-5)}\right),~~\nu=\pm\frac{4n+1}{2(4n-5)}.
\end{equation}
%%%
For $k\rightarrow 0$ limit, we have
%%%
\begin{eqnarray}
 \Delta&\propto&\tau^{\frac{2}{3}(2n-1)}=a^{\frac{2(2n-1)}{n+1}},
 ~~\tau^{-1}=a^{-\frac{3}{n+1}}, \\
 \Psi&\propto&\hbox{constant},~~\tau^{-\frac{1}{3}(4n+1)}=\frac{H}{a}.
\end{eqnarray}
%%%

%%%%%%%%%%%%%%%%%%%%%%%%%%%%%%%%%%%%%%%%%%%%%%%%%%%%%%%%%%%%%%%%%%%% 
%\section{summary}
%\begin{itemize}
%	\item  Evolution equation for $Q$ can be reduced to  Mathieu equation
%
%	\item  Growth of $\Delta$ is related to the unstable band of Mathieu 
%	function(parametric resonance)
%
%	\item  We does not assume ``Newtonian approximation''. We can treat 
%	the mode whose wave length is longer than ``Compton'' wave length
%
%	\item  ``Jeans'' scale appears
%	
%	\item extension to 2-scalar field system: straightforward.
%	
%	\item non-linear coupling between 2 scalar fields ???
%\end{itemize}
In summary, we found that the evolution equation of the Mukhanov's 
gauge invariant variable in the oscillatory phase of the scalar 
field can be reduced to the Mathieu equation and 
the evolution of this variable undergoes the effect of the parametric 
resonance. We can interpret the growth of the density perturbation in 
this phase is caused by the parametric resonance. 
Now we comment on previous works. In paper \cite{ns}, the analysis is 
done by using the Newtonian approximation which means the wavelength 
of the perturbation is smaller than the horizon length. But the 
obtained equation for $\left(\delta\rho/\rho\right)$ coincides with 
the result of this paper(eq.(\ref{u})). In paper \cite{kh}, the long 
wave approximation is used. As the eq.(\ref{muk}) has the exact 
solution for $k=0$, they take in the effect of small $k$ perturbatively 
and derive the evolution of the gauge invariant variables whose 
wavelength is larger than the horizon length. 
The assumption we used in this paper is the wavelength is larger than 
the Compton length which is smaller than the horizon scale in the 
oscillatory phase of the scalar field.
 So our treatment is 
more general. Extension to the non-linearly 
interacting two scalar field system which is a realistic model of the 
reheating is straightforward and the analysis is now going on. 
We will show the result in a separate publication.
%%%%%%% figure %%%%%%%%%%%%%%%%%
\newpage
\begin{center}
{\Large FIGURE}
\end{center}
The stability/instability chart of the Mathieu equation 
(\ref{math}). The grey is stable and the white is unstable 
region. Three curves shows the time evolution of the parameter $A, q$ 
for the power index $n$ of the scalar field potential 
$V=\frac{\lambda}{4}\phi_{0}^{4}\left(\frac{\phi}{\phi_{0}}\right)^{2n}$. 
$A=1$ line corresponds to $k=0$.
\newpage
\thispagestyle{empty}
\vspace*{22 cm}
%\begin{center}
%\epsfile{file=fig_reheat.eps,scale=1.0}
%\end{center}
\special{epsf=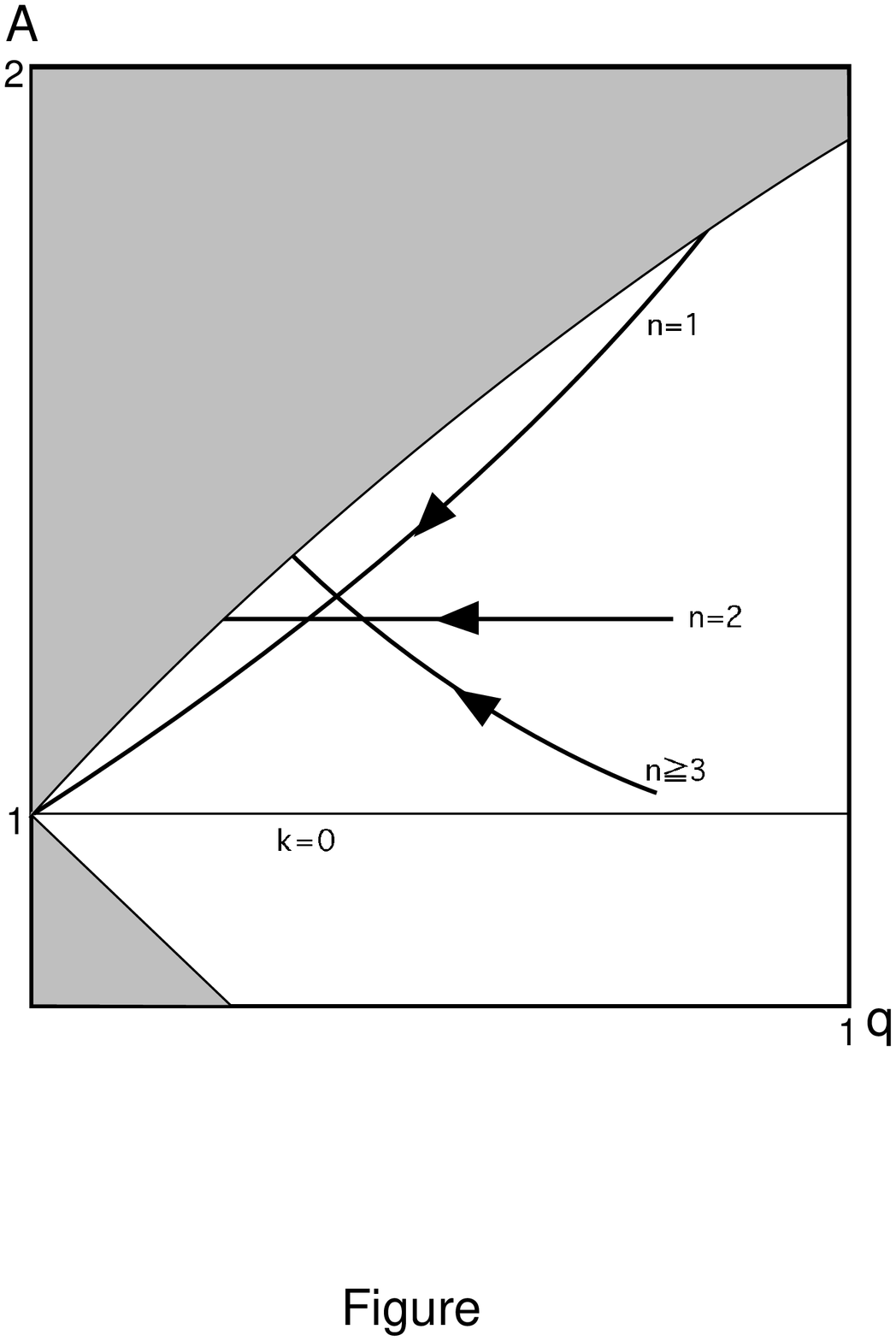}
%%%%%%%%%%%%%%%%%%%%%%%%%%%%%%%%%%%%%%%%%%%%%


\begin{references}
\bibitem{sht} Y. Shtanov, J. Traschen and R. Brandenberger, {\it Phys. Rev.} {\bf 
D51}(1995)5438.
\bibitem{kof} L. Kofman, A. Linde and A. A. Starobinsky, {\it Phys. Rev. 
Lett.} {\bf 73}(1994)1425.
\bibitem{ks} H. Kodama and M. Sasaki, {\it Prog. Theor. Phys. 
Suppl.}{\bf 78}(1991)103.
\bibitem{muk} V. F. Mukhanov, {\it JETP} {\bf 68}(1988)1297. 
\bibitem{nay} A. H. Nayfeh, {\it Introduction to Perturbation 
Techniques}(A Wiley-Interscience Publication, 1993).
\bibitem{ns}Y. Nambu and M. Sasaki, {\it Phys. Rev.}{\bf 
D42}(1990)3918.
\bibitem{kh}H. Kodama and T. Hamazaki, {\it preprint}~gr-qc/9608022.
\end{references}
\end{document}